\documentclass[fleqn,usenatbib]{mnras}

\usepackage{newtxtext,newtxmath}

\usepackage[T1]{fontenc}

\usepackage{chemformula} 
\usepackage[T1]{fontenc} 
\usepackage{multirow}
\usepackage{bigdelim}
\usepackage{longtable}
\usepackage{graphicx}
\usepackage{times}
\usepackage{amssymb}
\usepackage{amsmath}
\usepackage[utf8]{inputenc}
\usepackage[T1]{fontenc}

\title[FAUST I. The Class I hot corino L1551 IRS5]{FAUST I. The hot corino at the heart of the prototypical Class I protostar L1551 IRS5}

\author[E. Bianchi et al.]{
E. Bianchi,$^{1}$\thanks{E-mail: eleonora.bianchi@univ-grenoble-alpes.fr}
C. J. Chandler,$^{2}$
C. Ceccarelli,$^{1}$
C. Codella,$^{3,1}$
N. Sakai,$^{4}$
A. L\'{o}pez-Sepulcre,$^{1,5}$
\newauthor
L. T. Maud,$^{6}$
G. Moellenbrock,$^{2}$
B. Svoboda,$^{2}$
Y. Watanabe,$^{7}$
T. Sakai,$^{8}$
F. Ménard,$^{1}$
Y. Aikawa,$^{9}$
\newauthor
F. Alves,$^{10}$
N. Balucani,$^{11}$
M. Bouvier,$^{1}$
P. Caselli,$^{12}$
E. Caux,$^{13}$
S. Charnley,$^{14}$
S. Choudhury,$^{10}$
\newauthor
M. De Simone,$^{1}$
F. Dulieu,$^{15}$
A. Durán,$^{16}$
L. Evans,$^{13,3}$
C. Favre,$^{1}$
D. Fedele,$^{3}$
S. Feng,$^{17,18,19}$
\newauthor
F. Fontani,$^{3}$
L. Francis,$^{20,21}$
T. Hama,$^{22,23}$
T. Hanawa,$^{24}$
E. Herbst,$^{25}$
T. Hirota,$^{18}$
M. Imai,$^{26}$
\newauthor
A. Isella,$^{27}$
I. Jim\'enez-Serra,$^{28}$
D. Johnstone,$^{20,21}$
C. Kahane,$^{1}$
B. Lefloch,$^{1}$
L. Loinard,$^{16,29}$
\newauthor
M.J. Maureira,$^{10}$
S. Mercimek,$^{3,30}$
A. Miotello,$^{6}$
S. Mori,$^{9}$
R. Nakatani,$^{4}$
H. Nomura,$^{31}$
\newauthor
Y. Oba,$^{33}$
S. Ohashi,$^{4}$
Y. Okoda,$^{26}$
J. Ospina-Zamudio,$^{1}$
Y. Oya,$^{26}$
J. Pineda,$^{10}$
L. Podio,$^{3}$
\newauthor
A. Rimola,$^{33}$
D. Segura Cox,$^{12}$
Y. Shirley,$^{34}$
V. Taquet,$^{3}$
L. Testi,$^{6}$
C. Vastel,$^{13}$
S. Viti,$^{35}$
\newauthor
N. Watanabe,$^{32}$
A. Witzel,$^{1}$
C. Xue,$^{25}$
Y. Zhang,$^{4}$
B. Zhao,$^{10}$
and S. Yamamoto$^{26}$
\\
\\
$^{1}$Univ. Grenoble Alpes, CNRS, IPAG, 38000 Grenoble, France\\
$^{2}$National Radio Astronomy Observatory, PO Box O, Socorro, NM 87801, USA\\
$^{3}$INAF, Osservatorio Astrofisico di Arcetri, Largo E. Fermi 5, I-50125, Firenze, Italy\\
$^{4}$RIKEN Cluster for Pioneering Research, 2-1, Hirosawa, Wako-shi, Saitama 351-0198, Japan\\
$^{5}$Institut de Radioastronomie Millim\'{e}trique, 38406 Saint-Martin d’H\`{e}res, France\\
$^{6}$European Southern Observatory, Karl-Schwarzschild Str. 2, 85748 Garching bei München, Germany\\
$^{7}$Materials Science and Engineering, College of Engineering, Shibaura Institute of Technology, 3-7-5 Toyosu, Koto-ku, Tokyo 135-8548, Japan\\
$^{8}$Graduate School of Informatics and Engineering, The University of Electro-Communications, Chofu, Tokyo 182-8585, Japan\\
$^{9}$Department of Astronomy, The University of Tokyo, 7-3-1 Hongo, Bunkyo-ku, Tokyo 113-0033, Japan\\
$^{10}$Max-Planck-Institut für extraterrestrische Physik (MPE), Gießenbachstr. 1, D-85741 Garching, Germany\\
$^{11}$Department of Chemistry, Biology, and Biotechnology, The University of Perugia, Via Elce di Sotto 8, 06123 Perugia, Italy\\
$^{12}$Center for Astrochemical Studies, Max-Planck-Institut für extraterrestrische Physik (MPE), Gießenbachstr. 1, D-85741 Garching, Germany\\
$^{13}$IRAP, Université de Toulouse, CNRS, CNES, UPS, Toulouse, France\\
$^{14}$Astrochemistry Laboratory, Code 691, NASA Goddard Space Flight Center, 8800 Greenbelt Road, Greenbelt, MD 20771, USA\\
$^{15}$CY Cergy Paris Université, Sorbonne Université, Observatoire de Paris, PSL University, CNRS, LERMA, F-95000, Cergy, France\\
$^{16}$Instituto de Radioastronomía y Astrofísica , Universidad Nacional Autónoma de México, A.P. 3-72 (Xangari), 8701, Morelia, Mexico\\
$^{17}$CAS Key Laboratory of FAST, National Astronomical Observatory of China, Datun Road 20, Chaoyang, Beijing, 100012, P. R. China\\
$^{18}$National Astronomical Observatory of Japan, National Institutes of Natural Sciences, 2-21-1 Osawa, Mitaka, Tokyo 181-8588, Japan\\
$^{19}$Academia Sinica Institute of Astronomy and Astrophysics, No.1, Sec. 4, Roosevelt Rd, Taipei 10617, Taiwan, Republic of China\\
$^{20}$Department of Physics and Astronomy, University of Victoria, 3800 Finnerty Road, Elliot Building Victoria, BC, V8P 5C2, Canada\\
$^{21}$NRC Herzberg Astronomy and Astrophysics 5071 West Saanich Road, Victoria, BC, V9E 2E7, Canada\\
$^{22}$Komaba Institute for Science, The University of Tokyo, 3-8-1 Komaba, Meguro, Tokyo 153-8902, Japan\\
$^{23}$Department of Basic Science, The University of Tokyo, 3-8-1 Komaba, Meguro, Tokyo 153-8902, Japan\\
$^{24}$Center for Frontier Science, Chiba University, 1-33 Yayoi-cho, Inage-ku, Chiba 263-8522, Japan\\
$^{25}$Department of Chemistry, University of Virginia, McCormick Road, PO Box 400319, Charlottesville, VA 22904, USA\\
$^{26}$Department of Physics, The University of Tokyo, 7-3-1, Hongo, Bunkyo-ku, Tokyo 113-0033, Japan\\
$^{27}$Department of Physics and Astronomy, Rice University, 6100 Main Street, MS-108, Houston, TX 77005, USA\\
$^{28}$Centro de Astrobiologia (CSIC-INTA), Ctra. de Torrejon a Ajalvir, km 4, 28850, Torrejon de Ardoz, Spain\\
$^{29}$Instituto de Astronomía, Universidad Nacional Autónoma de México, Ciudad Universitaria, A.P. 70-264, Cuidad de México 04510, Mexico\\
$^{30}$Università degli Studi di Firenze, Dipartimento di Fisica e Astronomia, via G. Sansone 1, 50019 Sesto Fiorentino, Italy\\
$^{31}$Division of Science, National Astronomical Observatory of Japan, 2-21-1 Osawa, Mitaka, Tokyo 181-8588, Japan\\
$^{32}$Institute of Low Temperature Science, Hokkaido University, N19W8, Kita-ku, Sapporo, Hokkaido 060-0819, Japan\\
$^{33}$Departament de Química, Universitat Autònoma de Barcelona, 08193 Bellaterra, Spain\\
$^{34}$Steward Observatory, 933 N Cherry Ave., Tucson, AZ 85721 USA\\
$^{35}$Department of Physics and Astronomy, University College London, Gower Street, London, WC1E 6BT, UK\\
}

\date{Accepted XXX. Received YYY; in original form ZZZ}
\date{Accepted XXX. Received YYY; in original form ZZZ}

\pubyear{2015}

\begin{document}
\label{firstpage}
\pagerange{\pageref{firstpage}--\pageref{lastpage}}
\maketitle

\begin{abstract}
The study of hot corinos in Solar-like protostars has been so far mostly limited to the Class 0 phase, hampering our understanding of their origin and evolution. 
In addition, recent evidence suggests that planet formation starts already during Class I phase, which, therefore, represents a crucial step in the future planetary system chemical composition.
Hence, the study of hot corinos in Class I protostars has become of paramount importance.
Here we report the discovery of a hot corino towards the prototypical Class I protostar L1551 IRS5, obtained within the ALMA Large Program FAUST.
We detected several lines from methanol and its isopotologues ($^{13}$CH$_{\rm 3}$OH and CH$_{\rm 2}$DOH), methyl formate and ethanol. 
Lines are bright toward the north component of the IRS5 binary system, and a possible second hot corino may be associated with the south component. 
The methanol lines non-LTE analysis constrains the gas temperature ($\sim$100 K), density ($\geq$1.5$\times$10$^{8}$ cm$^{-3}$), and emitting size ($\sim$10 au in radius). 
All CH$_{\rm 3}$OH and $^{13}$CH$_{\rm 3}$OH lines are optically thick, preventing a reliable measure of the deuteration. 
The methyl formate and ethanol relative abundances are compatible with those measured in Class 0 hot corinos.
Thus, based on the present work, little chemical evolution from Class 0 to I hot corinos occurs.
\end{abstract}

\vspace{-0.5cm}
\section{Introduction} \label{sec:intro}
Solar-like planetary systems are the result of a complex process that starts from a cold molecular cloud and evolves through various phases \citep[e.g.,][]{Caselli2012}.
Among them, the Class I protostellar stage, whose typical duration is $\leq$ 10$^5$ yr, represents a crucial link between the youngest Class 0 and the Class II/III protostars \citep[e.g.][]{Crimier2010}, the latter being characterized by developed protoplanetary disks.
A recent ALMA breakthrough was the detection of gaps and rings in disks around protostars with ages $\leq$ 1 Myr (\citealt{Sheehan2017, Fedele2018}), strongly suggesting that the planet formation process starts already in Class I protostellar disks.
Since the process itself and the chemical content of the future planets, asteroids and comets depend on the chemical composition of the disk/envelope, understanding it at the planet-formation scales has become crucial.

However, despite its far reaching importance, the chemical content of Class I protostars is, at the moment, poorly known.
Class 0 protostars have infalling-rotating envelopes and circumstellar disks whose chemical composition largely, but not exclusively, depends on the distance from the central accreting object and the composition of the grain mantles \citep[e.g.,][]{Caselli2012,Sakai2013}.
Particularly relevant to this Letter, Class 0 protostars possess hot corinos \citep{Ceccarelli2004}, which are defined as warm ($\geq100$ K), dense ($\geq10^7$ cm$^{-3}$) and compact ($\leq 100$ au) regions enriched in interstellar Complex Organic Molecules (hereinafter iCOMs; \citealt{Ceccarelli2017}). 
The chemical composition in these regions is believed to be the result of the sublimation of the grain mantles where the dust reaches about 100 K, regardless of the detailed geometry of the region, whether a spherical infalling envelope or a circumstellar disk.
While about a dozen Class 0 hot corinos are imaged so far, only two Class I hot corinos are \citep{Desimone2017,Bergner2019,Belloche2020}.
More generally, few studies have focused on Class I protostars, often targeting the envelope or specific molecules \citep{Jorgensen2004,Codella2016,Codella2018,Bianchi2017a, Bianchi2019a,Bianchi2019b,Bergner2018,Bergner2019,Bergner2019b,Artur2019a,Artur2019b,Oya2019}.
The scarcity of available observations makes it difficult to assess whether or not the chemical composition of Class 0 and I protostars differs. 
Observations of the chemical content of Class I protostars at the planet-formation scale have now become urgent to understand the chemical evolution during the formation of planetary systems around Solar-like stars.

In this context, the ALMA (Atacama Large Millimeter/submillimeter Array) Large Program FAUST (Fifty AU STudy of the chemistry in the disk/envelope system of Solar-like protostars; {\it http://faust-alma.riken.jp}) is designed to survey the chemical composition of a sample of 13 Class 0/I protostars at the planet-formation scale, probing regions from about 1000 to 50 au (all have a distance $\leq$250 pc).
The selected sources represent the protostellar chemical diversity observed at large ($\geq1000$ au) scales.
All the targets are observed in three frequency setups chosen to study both continuum and line emission from specific molecules: 85.0--89.0 GHz, 97.0--101.0 GHz, 214.0--219.0 GHz, 229.0--234.0 GHz, 242.5--247.5 GHz, and 257.5--262.5 GHz.
The FAUST survey provides a uniform sample in terms of frequency setting, angular resolution and sensitivity.
We report the first results, obtained towards the prototypical Class I protostar L1551 IRS5. 
This study focuses on iCOM lines and aims at discovering and studying its hot corino(s).

\vspace{-0.5cm}
\section{The L1551 IRS5 source}\label{sec:source}
L1551 IRS5 is located in Taurus \citep{Strom1976} at a distance of (141$\pm$7) pc \citep{Zucker2019}, has a $L_{\rm bol}$ = 30--40 $L_{\odot}$ \citep{Liseau2005}, is a FU Ori-like object \citep{Connelley2018} and is considered a prototypical Class I source \citep{Adams1987,Looney1997}.
It is surrounded by a large ($\sim$ 0.1 pc) rotating/infalling envelope with $A_{\rm v}$ $>$ 100 mag \citep[e.g.,][]{Fridlund2005, White2006, Moriarty2006} studied via lines from several species such as CH$_{\rm 3}$OH, HCN, CS, and HCO$^+$ \citep[e.g.,][]{Fridlund2002, White2006}. 
L1551 IRS5 is associated with a molecular outflow and the HH154 jet, extensively studied in the X-ray, optical, near IR, and radio emission \citep[e.g.,][and references therein]{Snell1980, Fridlund2005, Schneider2011}. 
Zooming into the inner 100 au, L1551 IRS5 is a binary system, revealed for the first time by VLA cm-observations \citep{Bieging1985} that show two parallel jets \citep{Rodriguez2003}.
The binarity was confirmed by BIMA millimeter observations \citep{Looney1997} that identified a northern source N ($\sim$0.8 $M_{\odot}$) and a southern source S ($\sim$0.3 $M_{\odot}$) \citep{Liseau2005}. 
The two protostars are surrounded by a circumbinary disk, whose radius and mass are $\sim$140 au and 0.02-0.03 M$_{\odot}$ \citep{Looney1997,Cruz2019}.
ALMA observations also suggest the presence of two dusty disks (M$_{disk}$>0.006M$_\odot$) towards N and S, with radii between 8 and 14 au. The protostellar disks inclination is expected to be $\sim$ 35--45$^{\circ}$ for N and $\sim$ 24--44$^{\circ}$ for S \citep{Cruz2019,Lim2016}.
Proper motion measurements show an orbital rotation of N and S with a period of $\sim$260 yr and an eccentricity orbit tilted by up to 25$^{\circ}$ from the circumbinary disk \citep{Rodriguez2003,Lim2016}.

\vspace{-0.5cm}
\section{Observations} \label{sec:obs}
L1551 IRS5 was observed with ALMA (FAUST Large Program 2018.1.01205.L).
The data here exploited were acquired on 2018 October 25 using the C43-5 antenna configuration, with baselines between 15 m and 1.4 km. 
The analysed spectral window (232.8--234.7 GHz) was observed using spectral channels of 488 kHz (0.63 km s$^{-1}$).
The observations were centered at
$\alpha_{\rm J2000}$ = 04$^{\rm h}$ 31$^{\rm m}$ 34$\fs$14,
$\delta_{\rm J2000}$ = +18$\degr$ 08$\arcmin$ 05$\farcs$10.
The quasar J0423-0120 was used as bandpass and flux calibrator, J0510+1800 as phase calibrator. 
The data were calibrated using the ALMA calibration pipeline within CASA \citep{McMullin2007} and we included an additional calibration routine to correct for the T$_{\rm sys}$ and spectral data normalization\footnote{https://help.almascience.org/index.php?/Knowledgebase/Article/View/419; Moellenbrock et al. (in preparation).}. 
The data were self-calibrated using carefully-determined line-free continuum channels, including corrections for the continuum spectral index, and the continuum model was then subtracted from the visibilities prior to imaging the line data. The resulting continuum-subtracted line-cube, made using a Briggs robust parameter of 0.4, has a synthesized beam of 0$\farcs$37$\times$0$\farcs$31 (PA=39$^{\circ}$), and an r.m.s. noise of 1 mJy beam$^{-1}$ in an 0.6 km s$^{-1}$ channel, as expected. We estimate the absolute flux calibration uncertainty of 10\% and an additional error of 10\% for the spectra baseline determination. Spectral line imaging was performed with the CASA package, while the data analysis was performed using the IRAM-GILDAS package.

\begin{table}
	\caption{Properties of 
	the lines detected towards L1551 IRS5.}
	\label{Tab:lines}
	\begin{tabular}{lccccc} 
		\hline
Transition & $\nu$$^{\rm a}$ & E$_{\rm up}$$^{\rm a}$ & $S\mu^2$$^{\rm a}$  &I$_{\rm int}$$^{\rm b}$ \\
 & (GHz) & (K) & (D$^2$) & (K km s$^{-1}$)\\
\hline
\hline
CH$_{\rm 3}$OH 10$_{\rm 3,7}$--11$_{\rm 2,9}$ E & 232.9458 & 190 & 12 & 61   \\ 
CH$_{\rm 3}$OH 18$_{\rm 3,15}$--17$_{\rm 4,14}$ A & 233.7957 & 447 & 22 &  54  \\
CH$_{\rm 3}$OH 4$_{\rm 2,3}$--5$_{\rm 1,4}$ A & 234.6834 & 61 & 4 &   76 \\ 
CH$_{\rm 3}$OH 5$_{\rm 4,2}$--6$_{\rm 3,3}$ E & 234.6985 & 123 & 2 &49 \\ 
\hline
$^{13}$CH$_{\rm 3}$OH 5$_{\rm 1,5}$--4$_{\rm 1,4}$A  & 234.0116 & 48 &4 &  54 \\
\hline
CH$_{\rm 2}$DOH 3$_{\rm 3,1}$ e0--4$_{\rm 2,2}$ e0 & 232.9019 & 49 & 0.2 &  10 \\
CH$_{\rm 2}$DOH 3$_{\rm 3,0}$ e0--4$_{\rm 2,3}$ e0 & 232.9290 & 49 & 0.2 &  34 \\
CH$_{\rm 2}$DOH 5$_{\rm 3,2}$ o1--4$_{\rm 2,2}$ e0 & 233.0831 & 68  & 0.2 & 45 \\
CH$_{\rm 2}$DOH 5$_{\rm 3,3}$ e0 --4$_{\rm 2,3}$ o1 & 233.1339 & 68  & 0.2 &  30\\
CH$_{\rm 2}$DOH 14$_{\rm 2,12}$ o1 --14$_{\rm 1,13}$ o1 & 233.1418 & 261  & 6 &  33 \\
CH$_{\rm 2}$DOH 9$_{\rm 2,8}$ e1 --9$_{\rm 1,9}$ e1 & 233.4611 & 123 & 1.5 &  34 \\
CH$_{\rm 2}$DOH 8$_{\rm 2,6}$ e0 --8$_{\rm 1,7}$ e0 & 234.4710 & 94 & 10 &  59 \\
\hline
HCOOCH$_{\rm 3}$ 19$_{\rm 4,16}$--18$_{\rm 4,15}$ A & 233.2268 & 123 & 48 & 37\\
HCOOCH$_{\rm 3}$ 19$_{\rm 14,6}$--18$_{\rm 14,5}$ E & 233.4144 & 242 & 23 &  17  \\
HCOOCH$_{\rm 3}$ 19$_{\rm 12,8}$--18$_{\rm 12,7}$ E & 233.6710 & 208 & 30 &  17\\
HCOOCH$_{\rm 3}$ 18$_{\rm 4,14}$--17$_{\rm 4,13}$ E & 233.7540 & 114 & 46 &  47\\
HCOOCH$_{\rm 3}$ 18$_{\rm 4,14}$--17$_{\rm 4,13}$ A & 233.7775 & 114 & 46 &  36\\
HCOOCH$_{\rm 3}$ 19$_{\rm 11,8}$--18$_{\rm 11,7}$ E & 233.8452 & 192 & 34 & 31 \\
HCOOCH$_{\rm 3}$ 19$_{\rm 11,9}$--18$_{\rm 11,8}$ E & 233.8672 & 192 & 34 &  22 \\
HCOOCH$_{\rm 3}$ 19$_{\rm 10,10}$--18$_{\rm 10,9}$ E & 234.1346 & 178 & 37 &   30 \\
\hline
CH$_{\rm 3}$CH$_{\rm 2}$OH 14$_{\rm 5,9}$--14$_{\rm 4,10}$  & 232.9285 & 120 & 14 & 36\\
CH$_{\rm 3}$CH$_{\rm 2}$OH 13$_{\rm 5,8}$--13$_{\rm 4,9}$ & 233.5710 & 108 & 13 & 18\\
CH$_{\rm 3}$CH$_{\rm 2}$OH 13$_{\rm 5,9}$--13$_{\rm 4,10}$  & 233.9511 & 108 & 13 &  25\\
CH$_{\rm 3}$CH$_{\rm 2}$OH 12$_{\rm 5,8}$--12$_{\rm 4,9}$ & 234.2552 & 97 & 12 &   24\\
CH$_{\rm 3}$CH$_{\rm 2}$OH 10$_{\rm 5,5}$--10$_{\rm 4,6}$ & 234.6663 & 78 & 9 &  28\\
CH$_{\rm 3}$CH$_{\rm 2}$OH 10$_{\rm 5,6}$--10$_{\rm 4,7}$ & 234.7146 & 78 & 9 &  18\\
\hline
\end{tabular}\\
$^{\rm a}$ Spectroscopic parameters of CH$_{\rm 3}$OH and $^{13}$CH$_{\rm 3}$OH are from \citet{Xu1997} and \citet{Xu2008}, retrieved from the CDMS database \citep{Muller2005}. Those of CH$_{\rm 2}$DOH, HCOOCH$_{\rm 3}$ and {\it anti-}CH$_{\rm 3}$CH$_{\rm 2}$OH are from \citet{Pearson2008, Pearson2012} and \citet{Ilyushin2009},  retrieved from the JPL database \citep{Pickett1998}.
$^{\rm b}$ Integrated intensities (T$_{\rm B}$) derived at the position P3 (Fig. \ref{fig:Maps}). The associated errors are less than 1 K km s$^{-1}$. 
\end{table}
\vspace{-0.5cm}
\section{Results} \label{sec:results}
\begin{figure*}
\begin{center}
\includegraphics[scale=0.33]{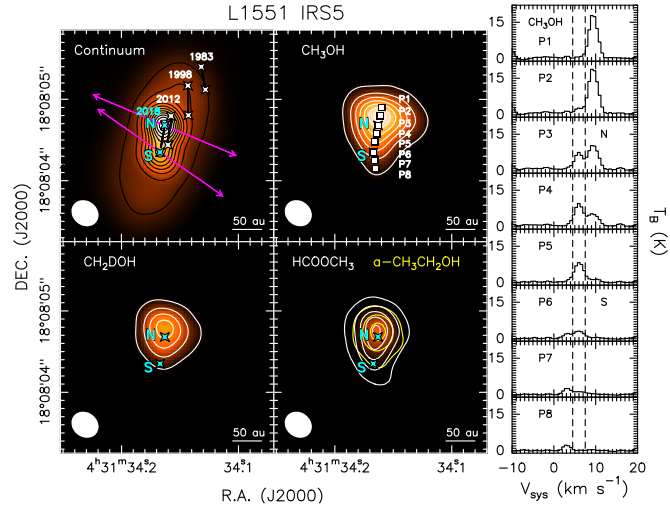}
\end{center}
\caption{Dust and line emission towards L1551 IRS5.
{\it Left Upper:} 1.3 mm dust continuum emission in colour scale and black contours. First contours and steps are 10$\sigma$ (1.8 mJy beam$^{-1}$) and 100$\sigma$, respectively.
The white stars indicate the positions of N and S measured in 1983, 1998 and 2012 \citep{Rodriguez2003b, Lim2016}, while the cyan stars refer to 2018 (this work). 
The magenta arrows indicate the jet directions \citep{Rodriguez2003}.
%
{\it Right Upper:}
Colour scale and white contours show the moment 0 map of the CH$_{\rm 3}$OH 5$_{\rm 4,2}$--6$_{\rm 3,3}$ E line, integrated over +2 to +15 km s$^{-1}$. First contours and steps are 8$\sigma$ (40 mJy beam$^{-1}$ km$^{-1}$) and 15$\sigma$, respectively.
The white squares, labelled from P1 to P8, are the different positions where the spectra displayed on the right panels are extracted. 
The positions P3 and P6 correspond to N and S, respectively.
{\it Left Lower}: 
Same as for the upper right panel, for the CH$_{\rm 2}$DOH 8$_{\rm 2,6}$--8$_{\rm 1,7}$ e0 line,  integrated over +2 to +15 km s$^{-1}$. First contours and steps are 8$\sigma$ (42 mJy beam$^{-1}$ km$^{-1}$) and 15$\sigma$, respectively.
{\it Right Lower}:
Colour scale and yellow contours show the moment 0 map of the a-CH$_{\rm 3}$CH$_{\rm 2}$OH 10$_{\rm 5,5}$--10$_{\rm 4,6}$ line integrated between +2 and +15 km s$^{-1}$. First contours and steps are 5$\sigma$ (42 mJy beam$^{-1}$ km$^{-1}$) and 10$\sigma$, respectively.
White contours show the moment 0 map of the HCOOCH$_{\rm 3}$ 18$_{\rm 4,14}$--17$_{\rm 4,13}$ A line integrated between +2 and +12 km s$^{-1}$. First contours and steps are 5$\sigma$ (28.5 mJy beam$^{-1}$ km$^{-1}$) and 10$\sigma$, respectively.
{\it Right panel}: CH$_{\rm 3}$OH 5$_{\rm 4,2}$--6$_{\rm 3,3}$ spectra extracted at the P1--P8 map positions.
The vertical dashed lines mark the systemic velocity inferred towards N (+7.5 km s$^{-1}$) and S (+4.5 km s$^{-1}$), respectively.
\label{fig:Maps}}
\end{figure*}

Figure \ref{fig:Maps} shows the map towards L1551 IRS5 of the dust emission at 1.3mm and the position of N and S at different dates since 1983. 
The two objects are not clearly resolved. The deconvolved source size, derived from a 2D Gaussian fit of the emission, is around 0$\farcs$4, similar to the beam size. In addition, it is evident that N is brighter, in agreement with previous observations \citep{Cruz2019}. 
The continuum map also shows extended emission ($\sim$1$\arcsec$ in radius) associated with the circumbinary disk.

\begin{figure*}
\begin{center}
\includegraphics[scale=0.4]{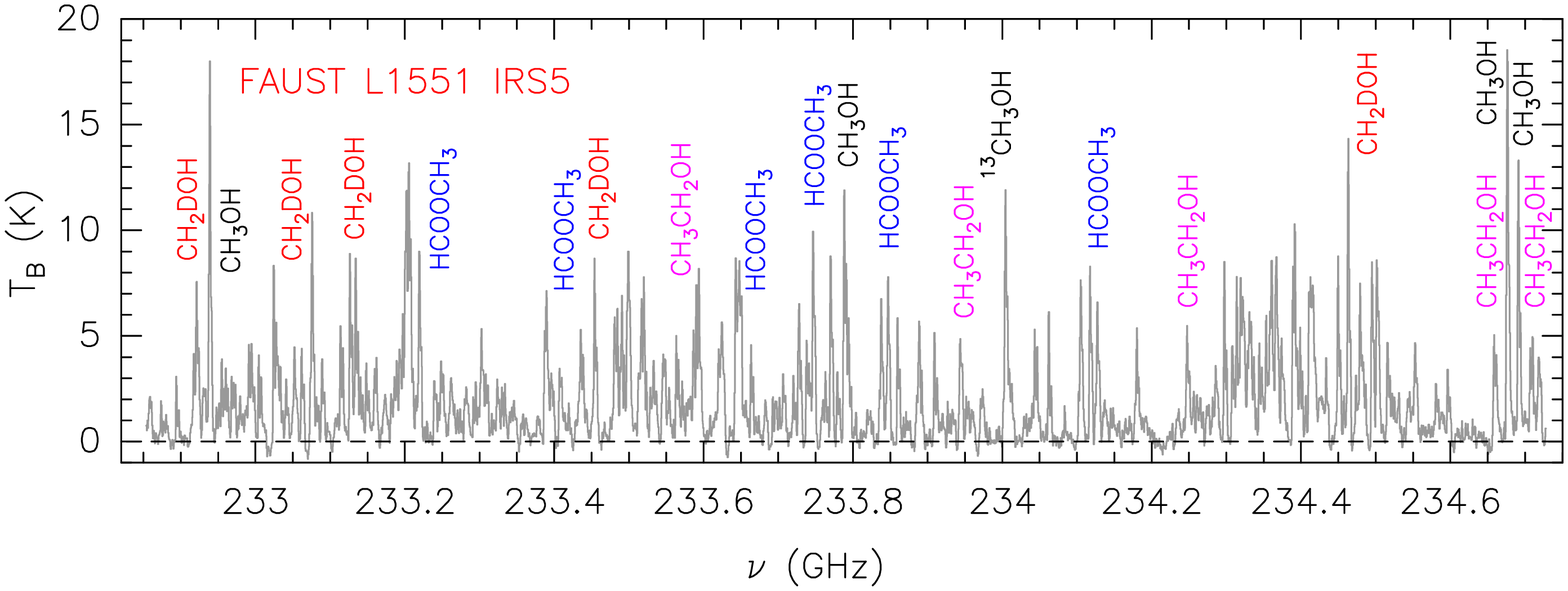}
\includegraphics[scale=0.31]{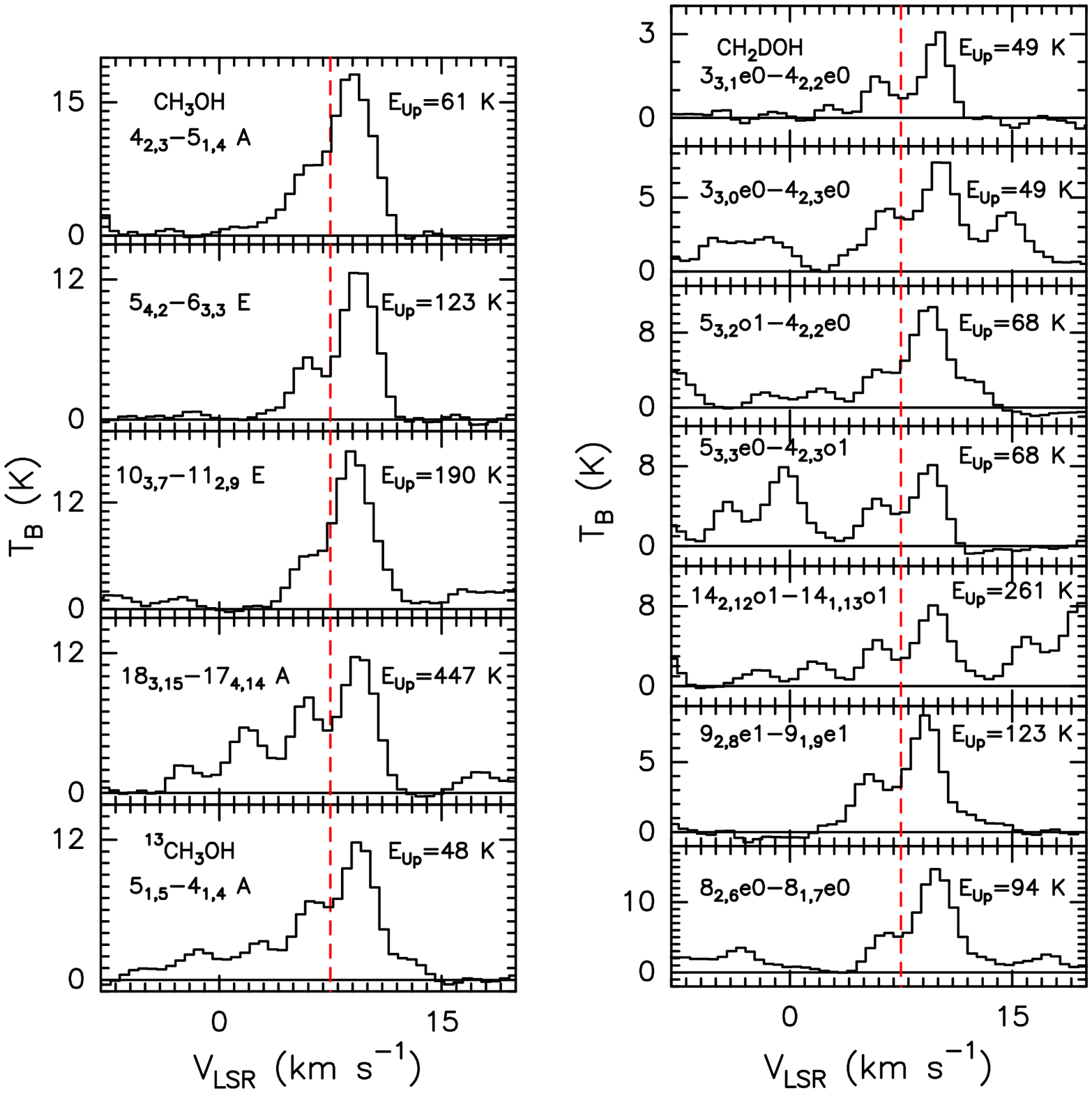}
\includegraphics[scale=0.31]{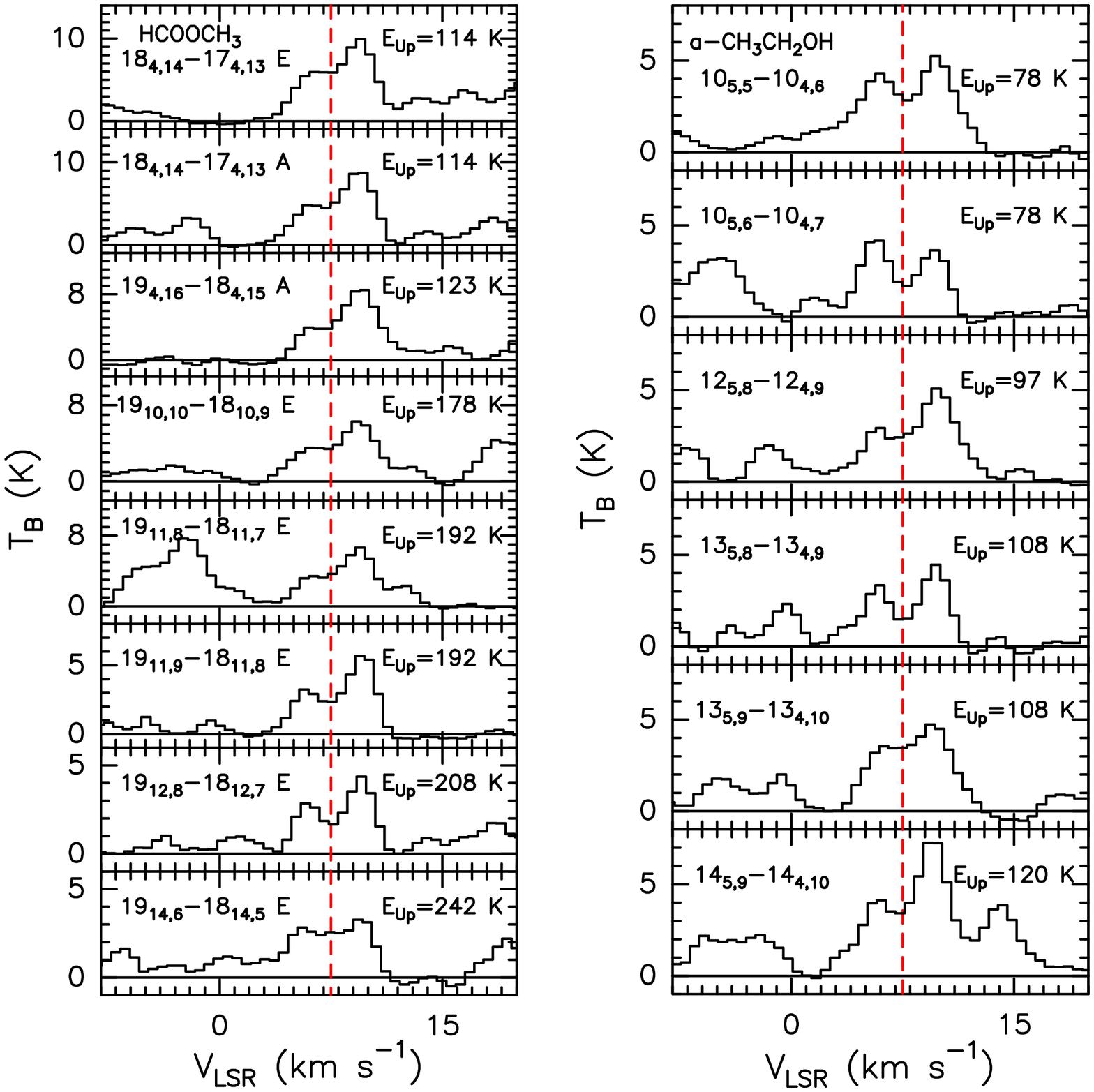}
\caption{Observed line spectra (in T$_{\rm B}$ scale) towards IRS5 N (P3 position of Fig. \ref{fig:Maps}). {\it Upper panel}: Entire spectrum between 232.8 and 234.7 GHz. {\it Lower panels}: Spectra of CH$_{\rm 3}$OH, $^{13}$CH$_{\rm 3}$OH, CH$_{\rm 2}$DOH, HCOOCH$_{\rm 3}$ and {\it anti-}CH$_{\rm 3}$CH$_{\rm 2}$OH: transitions and upper level energies are reported in the upper left and right corner of each panel, respectively. The vertical dashed line marks the ambient LSR velocity (+7.5 km s$^{-1}$).} \label{fig:spectra}
\end{center}
\end{figure*}

Table \ref{Tab:lines} lists the detected lines from the following iCOMs: methanol and its most abundant isotopologues (CH$_{\rm 3}$OH, $^{13}$CH$_{\rm 3}$OH, CH$_{\rm 2}$DOH), methyl formate (HCOOCH$_{\rm 3}$) and ethanol (CH$_{\rm 3}$CH$_{\rm 2}$OH).
In Fig. \ref{fig:Maps}, we show the integrated intensity (moment 0) maps for one representative line of each molecule.
For all lines, the emission peak coincides, within the synthesized beam, with the continuum position peak of source N, although fainter emission is also detected towards the southern component.
The figure also shows the spectra of the CH$_{\rm 3}$OH 5$_{\rm 4,2}$--6$_{\rm 3,3}$ E line, extracted in one pixel from different positions across the region in a direction perpendicular to the jet direction. 
Note that the spectra of the other iCOM lines have the same behavior. Specifically:

\noindent
{\it IRS5 N:} The methanol emission towards N, marked as P3 in Fig. \ref{fig:Maps}, has a double-peaked profile with a central dip at +7.5 km s$^{-1}$. 
The red- and blue-shifted peaks seem associated with gas to the north (positions P1 and P2) and south (P4 and P5) of N, respectively.
This velocity pattern, perpendicularly to the jet axis, could be due to either a rotating inner envelope or a disk. Unfortunately, since the emission is not resolved, it is impossible to discriminate between the two possibilities.
The spectra of all detected iCOM lines towards position P3, corresponding to the N continuum peak, are shown in Fig. \ref{fig:spectra}, while their spectral parameters are reported in Tab. \ref{Tab:lines}.

\noindent
{\it IRS5 S:} Similarly to N, the lines are double-peaked towards S, with a central dip at +4.5 km s$^{-1}$, namely $\sim 3$ km s$^{-1}$ red-shifted with respect to N. Going south (positions P7 and P8) the red peak disappears and only the blue one remains, suggesting again emission from a rotating inner envelope or a disk, assuming that the red peak is mainly associated with S.

\vspace{-0.5cm}
\section{Column densities and physical parameters}\label{sec-col-dens}

We derived the density and temperature of the gas emitting the methanol lines towards P3 (Tab. \ref{Tab:lines}), along with the molecular abundances of the detected iCOMs.
To this end, we carried out a non-LTE analysis of the CH$_{\rm 3}$OH lines via the Large Velocity Gradient (LVG) code by \citet{Ceccarelli2003}. 
We used the collisional coefficients of CH$_{\rm 3}$OH-A and CH$_{\rm 3}$OH-E with para-H$_{\rm 2}$ computed between 10 and 200 K for the first 256 levels of each species \citep{Rabli2010}, provided by the BASECOL database \citep{Dubernet2013}. 
We assumed a spherical geometry \citep{de_jong_hydrostatic_1980}, a CH$_3$OH-A/CH$_3$OH-E ratio equal to 1, a $^{12}$C/$^{13}$C ratio equal to 60 \citep{Milam2005}, a line FWHM equal to 3.5 km s$^{-1}$ as measured, and that the levels are populated by collisions and not by the absorption of the dust background photons whose contribution is very likely negligible due to the low values of the CH$_3$OH Einstein coefficients. We ran a large grid of models ($\geq10^4$) covering a CH$_{\rm 3}$OH-A (N$_{\rm CH_3OH-A}$) and CH$_{\rm 3}$OH-E column density from 3$\times$10$^{16}$ to 4$\times$10$^{19}$ cm$^{-2}$, an H$_{\rm 2}$ number density
(n$_{\rm H_2}$) from 3$\times$10$^{6}$ to 2$\times$10$^{8}$ cm$^{-3}$, and a gas temperature (T$_{\rm kin}$) from 80 to 180 K.
We found the solution with the lowest $\chi^2$ by simultaneously fitting the CH$_{\rm 3}$OH-A, CH$_{\rm 3}$OH-E and $^{13}$CH$_{\rm 3}$OH-A lines, leaving the N$_{\rm CH_3OH-A}$, n$_{\rm H_2}$, T$_{\rm kin}$, and the emitting size (the emission is unresolved) as free parameters. Since the collisional coefficients are available for only the three lines with the upper level energy less than 200 K, only those were used.
The best fit is obtained with N$_{\rm CH_3OH-A}$ = 1$\times$10$^{19}$ cm$^{-2}$ and a size of 0$\farcs$15 (21 au). Solutions with N$_{\rm CH_3OH-A}$ $\geq$ 0.5$\times$10$^{19}$ cm$^{-2}$ are within the 1$\sigma$ confidence level. All the observed CH$_{\rm 3}$OH lines are optically thick ($\tau >$ 50), as well as the $^{13}$CH$_{\rm 3}$OH-A ($\tau\sim2$) one, which makes the size well constrained.
The temperature is (100$\pm$10) K and the density is $\geq1.5\times$10$^{8}$ cm$^{-3}$ at 1$\sigma$ confidence level. The results do not change if we assume a line FWHM of 3.0 or 4.0 km s$^{-1}$ and they as the line optical depths are weakly model-dependent because of the $^{13}$CH$_{\rm 3}$OH line detection.

Collisional rates are not available for the other molecules, so we used the Rotational Diagram analysis to estimate their column densities, assuming a source size of 0$\farcs$15 as derived from the methanol analysis. 
In the case of CH$_{\rm 2}$DOH, we derive a rotational temperature of 88$\pm$9 K and a column density (64$\pm$11)$\times10^{16}$ cm$^{-2}$. 
However, as the non-LTE methanol line analysis shows that even the $^{13}$CH$_{\rm 3}$OH line is optically thick, we expect the same for the CH$_{\rm 2}$DOH lines, so that the derived column density is a lower limit.
For HCOOCH$_{\rm 3}$ and CH$_{\rm 3}$CH$_{\rm 2}$OH, the E$_{\rm up}$ range covered by the detected lines is not large enough, so we assumed a rotational temperature of 100 K, based on the methanol LVG analysis, to derive the respective column densities. 
They are (33$\pm$2)$\times10^{16}$ and (149$\pm$13)$\times10^{15}$ cm$^{-2}$, for methyl formate and ethanol, respectively.
With these column densities, the predicted opacity is around 0.3--0.5 for the methyl formate lines and $\sim$0.2 for the ethanol lines.
Therefore, both column densities (Table \ref{Table:detected}) are not affected by the line opacity.

\vspace{-0.5cm}
\section{Discussion and conclusions} \label{sec:discussion}
\subsection{The hot corinos of L1551 IRS5}\label{sec:hot-corinos}
The derived gas temperature and the detection of iCOMs make L1551 IRS5 N a hot corino.
The present data also suggest the presence of a second hot corino in S, to be confirmed by higher spatial resolution observations.
This increases, and perhaps doubles, the number of known Class I hot corinos as, before this work, only two were imaged, SVS13-A \citep{Desimone2017, Belloche2020} and Ser-emb 17 \citep{Bergner2019}.
Besides, our observations are the first to provide the chemical richness of Class I protostars on a Solar System scale. 
The derived emitting size for N of 0$\farcs$15, equivalent to about 20 au, is consistent with the heating from the central 40 L$_\odot$ source, and does not necessarily require an outburst activity.
However, note that the 0$\farcs$15 sizes are derived assuming a filling factor from a circular gaussian source emission. If the emission is more elongated in one direction, as would be the case in a rotating envelope and/or disk, this could explain the slightly more extended emission of Fig. \ref{fig:Maps}.

One result of the present work is that the methanol lines towards L1551 IRS5 N are very optically thick.
This implies that we can only establish a lower limit to the true methanol column density. 
This large methanol line opacity very likely is not a unique property of L1551 IRS5 and it is even more dramatic in Class 0 protostars, with their larger material column densities with respect to Class I sources.
This was already clear from the observations of IRAS16293--2422, where CH$_{\rm 3}^{18}$OH was used to derive the methanol column density \citep{Jorgensen2016}. Even more dramatically, recent VLA observations showed extremely optically thick methanol lines towards NGC1333 IRAS4A1 and IRAS4A2 \citep{Desimone2020}.
Here we show that even in Class I hot corinos the estimation of the column density of methanol assuming that the $^{13}$C isotopologue lines are optically thin can be inaccurate.
This fact could explain the contradictory results found by \citet{Bianchi2019a} when comparing the iCOM abundances normalised to methanol in different Class 0 and I protostars. 
A reliable measure requires the $^{18}$O methanol isotopologue detection.

Finally, given the high line optical depths, we can not estimate the methanol deuteration, because both the derived methanol and deuterated methanol column densities are lower limits, $\geq 1\times$ 10$^{19}$ cm$^{-2}$ and $\geq 5\times$ 10$^{17}$ cm$^{-2}$, respectively.
Taking these at face value, methanol deuteration would be of 5\%.
Again, to obtain a reliable measure requires the detection of $^{13}$CH$_2$DOH.

\begin{table}
\caption{List of detected iCOMs towards L1551 IRS5 N.}
\begin{center}

\begin{tabular}{lccrr}
  \hline
  \multicolumn{1}{c}{Species} &
                                \multicolumn{1}{c}{N$_{lines}^a$} &
                                                                  \multicolumn{1}{c}{E$_{\rm u}$} &
                                                                                                    \multicolumn{1}{c}{T$_{\rm rot}^b$} &
                                                                                                                                            \multicolumn{1}{c}{N$_{\rm tot}^b$}\\
  \multicolumn{1}{c}{} &
                         \multicolumn{1}{c}{} &
                                                \multicolumn{1}{c}{(K)} &
                                                                          \multicolumn{1}{c}{(K)} &
                                                                                                    \multicolumn{1}{c}{(cm$^{-2}$)}\\
                                                                                                
  \hline
    \multicolumn{5}{c}{non-LTE analysis}\\
CH$_{\rm 3}$OH  & 3 & 61--190 & 100(10) & $\geq 1\times 10^{19}$$^c$ \\
  
  $^{13}$CH$_3$OH & 1 & 48 &  &  \\
  
  \hline
      \multicolumn{5}{c}{Rotational Diagram analysis}\\

  CH$_{\rm 2}$DOH  & 7 & 49--261 & 88(9) & $\geq 5\times 10^{17}$\\
    
  HCOOCH$_{\rm 3}$  &  8 & 114--242 &   100$^d$ &  33(2) $\times 10^{16}$$^d$ \\
  
 CH$_{\rm 3}$CH$_{\rm 2}$OH  &  6 & 78--120 &   100$^d$ &  149(13) $\times 10^{15}$$^d$ \\

 \hline
\end{tabular}\\
$^a$ Number of lines used in the analysis.
$^b$ Parameters derived adopting a source size of 0$\farcs$15, as derived by the non-LTE analysis of the methanol lines. Upper limits and error bars (in parenthesis) are at 1$\sigma$ confidence level.
$^c$ Total methanol column density.
$^d$ To derive the column density we assumed $T_{\rm rot}$ equal to 100 K, as derived by the methanol non-LTE analysis.
 \end{center}
\label{Table:detected}
\end{table}

\vspace{-0.5cm}
\subsection{Methyl formate and ethanol in Class 0 and I sources}
The methyl formate and ethanol abundances relative to methanol are $\leq 0.03$ and $\leq 0.015$, respectively (Tab. \ref{Table:detected}).
The methyl formate normalised abundance is compatible with what measured, at comparable spatial scales, towards the Class 0 hot corinos IRAS 16293-2122B \citep[0.03;][]{Jorgensen2018}, HH212 \citep[0.03;][]{Lee2019}, IRAS 4A and IRAS 2A \citep[0.005 and 0.016;][]{Taquet2015,Lopez2017}. 
The ethanol normalised abundance in L1551 IRS5 N is also similar to the normalized abundances measured in the Class 0 hot corinos  mentioned above, namely 0.006--0.02.
Finally, both methyl formate and ethanol normalised abundances are similar to those measured in the Class I hot corino of SVS13-A, 0.016 and 0.014, respectively \citep{Bianchi2019a}.
A more reliable comparison can be obtained by considering the abundance ratio between methyl formate and ethanol, which are both optically thin. 
In L1551 IRS5 N, this value is $\sim$2, a factor 2 larger than that measured in the Class 0 IRAS16293-2122B \citep{Jorgensen2018} and Class I SVS13-A \citep{Bianchi2019a}.
Considering all the uncertainties, the Class I L1551 IRS5, similarly to SVS13-A, does not look dramatically different from Class 0 hot corinos with respect to the iCOM relative abundances.

\vspace{-0.5cm}
\section*{Acknowledgements}
This project has received funding from : 1) the European Research Council (ERC) under the European Union's Horizon 2020 research and innovation program, for the Project “The Dawn of Organic Chemistry” (DOC), grant agreement No 741002; 2) the PRIN-INAF 2016 The Cradle of Life - GENESIS-SKA (General Conditions in Early Planetary Systems for the rise of life with SKA); 3) a Grant-in-Aid from Japan Society for the Promotion of Science (KAKENHI: Nos. 18H05222, 19H05069, 19K14753); 4) the Spanish FEDER under project number ESP2017-86582-C4-1-R; 5) DGAPA, UNAM grants IN112417 and IN112820, and CONACyT, Mexico; 6) ANR of France under contract number ANR-16-CE31-0013; 7) the French National Research Agency in the framework of the Investissements d’Avenir program (ANR-15-IDEX-02), through the funding of the "Origin of Life" project of the Univ. Grenoble-Alpes, 8) the European Union’s Horizon 2020 research and innovation programs under projects “Astro-Chemistry Origins” (ACO), Grant No 811312.
This paper makes use of the following ALMA data: ADS/JAO.ALMA\#2018.1.01205.L. ALMA is a partnership of ESO (representing its member states), NSF (USA) and NINS (Japan), together with NRC (Canada), MOST and ASIAA (Taiwan), and KASI (Republic of Korea), in cooperation with the Republic of Chile. The Joint ALMA Observatory is operated by ESO, AUI/NRAO and NAOJ. The National Radio Astronomy Observatory is a facility of the National Science Foundation operated under cooperative agreement by Associated Universities, Inc. We thank the referee, Paul Ho, for his insightful suggestions.\\
{\bf DATA AVAILABILITY:} The raw data will be available on the ALMA archive at the end of the proprietary period (ADS/JAO.ALMA\#2018.1.01205.L).
%

\vspace{-0.5cm}
\bibliographystyle{mnras}
\bibliography{L1551} 




\bsp	
\label{lastpage}
\end{document}